\documentclass[]{aa} 
\usepackage{natbib}
\usepackage{graphicx}
\usepackage{subfigure}
\usepackage{txfonts}
\begin{document}
   \title{The high velocity outflow in NGC6334~I}
   \author{S. Leurini, P. Schilke, B. Parise,  F. Wyrowski, R. G\"usten \& S. Philipp}
   \offprints{S. Leurini}
   \institute{Max-Planck-Institut f\"ur Radioastronomie, Auf Dem H\"ugel 69,
     D-53121, Bonn \\
     \email{[sleurini,schilke,bparise,wyrowski,rguesten,sphilipp]@mpifr-bonn.mpg.de}}
   \date{}

  \abstract
   {}
   {We observed the high velocity outflow originating from NGC6334~I in several
     CO transitions with the APEX telescope, with the goal of deriving the 
physical parameters of the gas.}
   {Using an LVG analysis, we studied line ratios between the CO(3-2),
   CO(4-3), and CO(7-6) data as a function of the density and of the kinetic temperature of the gas. An upper 
limit on the CO column density is derived by comparison with $^{13}$CO data.}
   {We constrained the 
temperature to be higher than 50~K and the H$_2$ density to values higher
than $n\sim 10^4$~cm$^{-3}$ towards the peak
position in the red lobe, while $T>15$~K and $n>10^3$~cm$^{-3}$ are derived towards the peak
position in the blue lobe. The physical parameters of the outflow, its mass
and its energetics, have been computed using the temperatures derived from this
analysis.}
   {We conclude that high kinetic
temperatures are present in the outflow and traced by high excitation CO lines. Observations of
high-J CO lines are thus needed to infer reliable
values of the kinetic temperatures and of the other physical parameters in outflows.}
   \titlerunning{The high velocity outflow in NGC6334~I}
   \authorrunning{Leurini et al.}
\maketitle
%


\section{Introduction}
Despite molecular outflows being a well-studied phenomenon associated with young stellar 
objects of all masses, many questions on their driving mechanisms and
on their interaction with the surrounding
material are still unanswered. The case of massive outflows, 
which only recently have been investigated at high spatial resolution \citep[e.g.][]{2002A&A...387..931B,2004ApJ...616L..35S}, is even more problematic.
Usually, low rotational (J) CO transitions are used to trace molecular outflows and derive their physical parameters
and  energetics. However, these lines trace only cold gas, while high-J transitions \citep[e.g.][]{2002A&A...387..931B}
have also been detected, clearly pointing to hot, dense 
gas associated with the outflow.  High resolution observations of high-J CO transitions
are needed to get a better understanding of the phenomenon.

The region \object{NGC6334}, located at a distance of 1.7~kpc from the
Sun \citep{1978A&A....69...51N}, harbours a series of luminous star-forming regions at various stages of evolution.  NGC6334~I is the
dominant source at (sub)millimeter wavelengths
\citep{2000A&A...358..242S}, with a compact H{\sc ii} region at its
center \citep{1997ApJ...486L.103C}.  Sub-mm line surveys
(\citealt{1994MNRAS.271...75S,2000MNRAS.316..152M,2003ASPC..287..257T}, Schilke et al. this volume)
reveal rich molecular line emission from a dense, hot core of size
$\le 10''$.  Two outflows originate from it
\citep[][]{1990A&A...239..276B,2000MNRAS.316..152M}.

In this Letter, we present high frequency $^{13}$CO and  CO observations
of the main outflow originating from  NGC6334~I. The physical parameters of the outflow
are derived by a Large Velocity Gradient (LVG) analysis of several CO transitions. 
\section{Observations}
The observations of NGC6334~I ($\alpha_{\rm J2000}=$17:20:53.35,
$\delta_{\rm J2000} =$--35:47:01.50) were performed with the Atacama Pathfinder Experiment 12~m 
telescope
(APEX\footnote{This publication is based on data acquired with the Atacama
Pathfinder Experiment (APEX). APEX is a collaboration between the Max-Planck-Institut
  f\"ur Radioastronomie, the European Southern Observatory, and the Onsala Space Observatory.}, G\"usten et al. 2006, this volume)  during June, September, and November, 2005. The APEX-2A (Risacher et al. 2006, this volume) and
FLASH (Heyminck et al. 2006, this volume)  receivers were used to map the source in CO(3-2) and  CO(4-3). 
The second channel of the FLASH receiver was tuned to CO(7-6); 
we performed long
integration observations of the two positions where the blue and
redshifted emission in CO(4-3) is strongest, at the offset positions $(-10'',-14'')$ and
$(17.5'',14'')$.
  Additional observations of the same two positions in $^{13}$CO(3-2) were performed in
November 2005 with the APEX-2A receiver.  
All observations were taken in the lower side band.
For the CO(4-3) and CO(7-6)
observations, the system temperatures are around 800~K and 3400~K, at
461~GHz and 807~GHz respectively, while the system temperatures of the
APEX-2A receiver are of the order of 260~K for CO(3-2) and 480~K for
$^{13}$CO(3-2). The system temperatures of the $^{13}$CO(3-2) are higher,  as 
they were  performed in November 2005 when the source was already in the
Sun avoidance limit and was observed at low declination ($<20 ^\circ$)
after sunset.
 Pointing  was checked on the
continuum of NGC6334~I itself, during the CO(4-3) observations, and was found to be accurate to $\sim 5''$. On
the other hand, problems with the pointing were encountered during the
CO(3-2) run and the data show a systematic shift of $\sim8''$ with respect to the CO(4-3)
observations. Alignment of the CO(3-2) map to the CO(4-3) is possible since
several molecules are detected in the two bands that have high excitation
conditions and peak on the hot core. We estimate the accuracy to be good to 2-3", since it was based on the alignment of hot core lines, which trace the (within our beam) point like hot core.

The FFTS spectrometer (Klein et al. 2006, this volume) was used with resolutions of 
0.1~${\rm km~s}^{-1}$ for CO(3-2),
0.3~${\rm km~s}^{-1}$ for CO(4-3) and CO(7-6), and 0.4~${\rm km~s}^{-1}$ for $^{13}$CO(3-2).
The calibration was performed by using the APECS software (Muders et al. 2006, this volume).
 Antenna temperatures were converted to main-beam temperatures
by using forward and beam efficiencies 
of 0.97 and 0.74 at 345~GHz, 0.95 and 0.60
at 464~GHz, and 0.95  and 0.43 at 809~GHz.

\begin{figure}[h]
  \includegraphics[angle=-90,width=7cm]{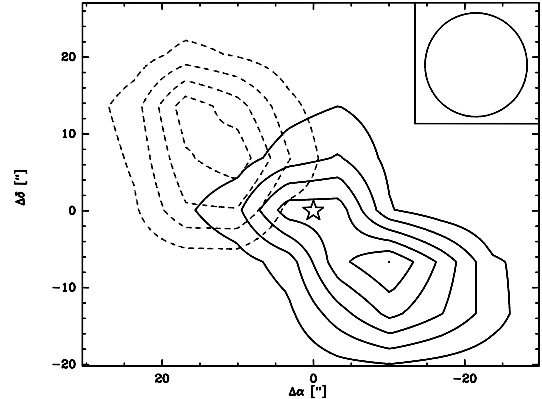}
\caption{Map of the CO(4-3) emission in the $7<v<55$~km/s (red, dashed line) and
$-78<v<-15$~km/s (blue, solid line) ranges. Contours are from 100 to
600~K~km/s by steps of 100. The star marks the NGC6334~I position; the beam ($\sim 13.5''$) is indicated in the  top right corner.}\label{map}
\end{figure}

\begin{figure}
\includegraphics[angle=-90,width=7cm]{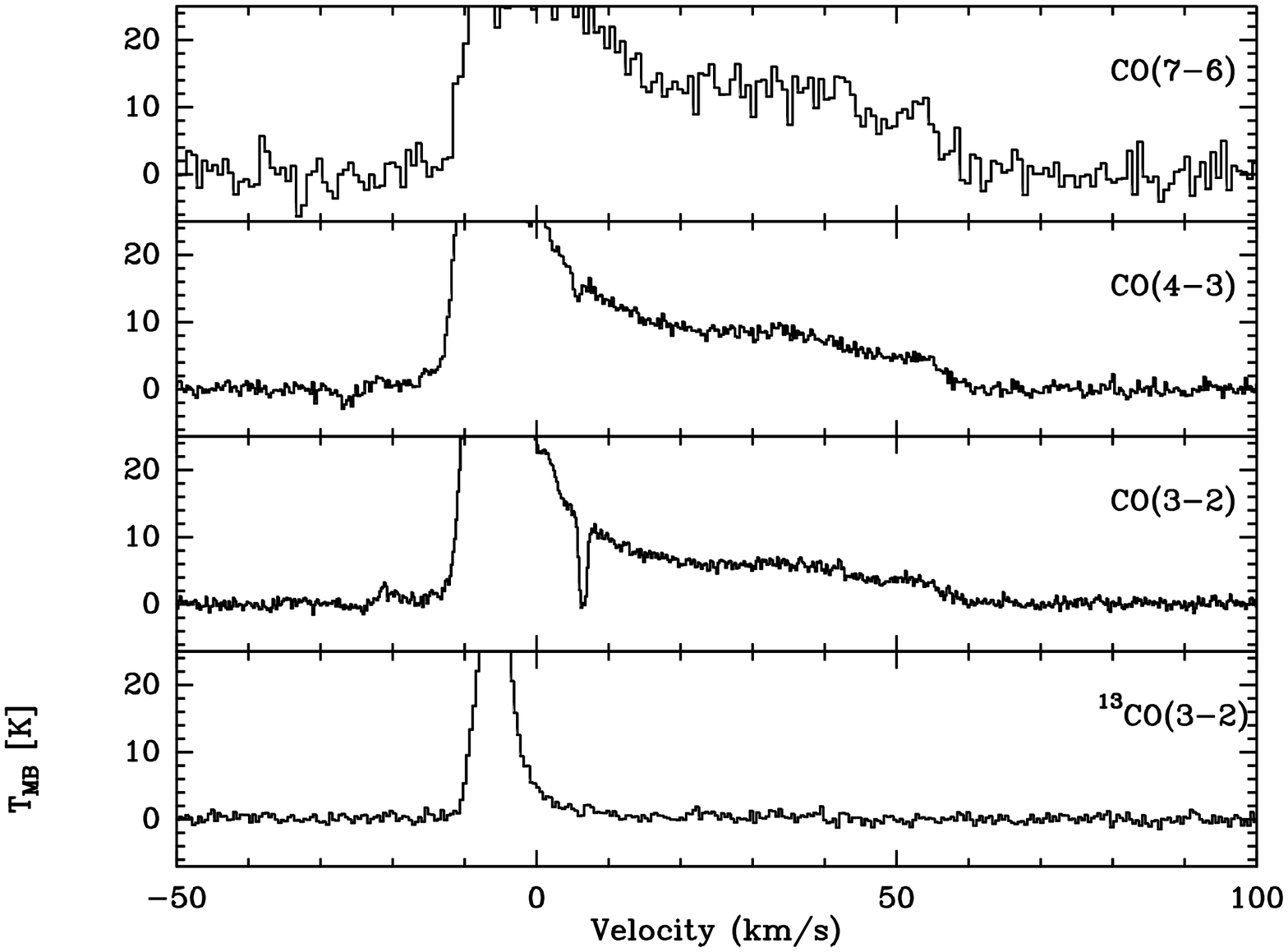}
\caption{CO(J=3,4,7)  and $^{13}$CO(3-2) spectra at the offset $(17.5'',14'')$ from the reference position ($\alpha_{\rm J2000}=$17:20:53.35, $\delta_{\rm J2000}=-35:47:01.50$). The CO(7-6) and CO(4-3) spectra are smoothed to the CO(3-2) spatial resolution.}\label{red}
\end{figure}

\begin{figure}
\includegraphics[angle=-90,width=7cm]{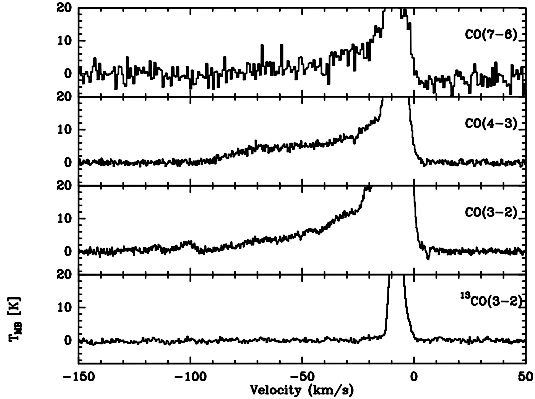}
\caption{CO(J=3,4,7)  and $^{13}$CO(3-2) spectra at the offset $(-10'',-14'')$ from the reference position ($\alpha_{\rm J2000}=$17:20:53.35, $\delta_{\rm J2000}=-35:47:01.50$). The CO(7-6) and CO(4-3)
  spectra are smoothed to the CO(3-2) spatial resolution.}\label{blue}
\end{figure}

The CO(4-3) map is shown in Fig.~\ref{map}, while spectra towards the $(-10'',-14'')$ and
$(17.5'',14'')$ positions are presented in Figs.~\ref{red}-\ref{blue}.   
\citet{1990A&A...239..276B} detected emission at --100~km~s$^{-1}$ in the
CO(2-1) spectra; they did  not identify it with a CO velocity component,
even though  they could
not assign the feature to any other molecule, because it
peaks at the hot core position.
The detection of a similar feature at --100~${\rm km~s}^{-1}$ in our CO(3-2) spectra 
(Fig.~\ref{blue}) supports the interpretation that this is indeed another CO
high velocity component. 
 Alternatively, the feature can be identified as CH$_3$OH
($16_1\to15_2$, E$_{\rm{low}}\sim316$~K) as it peaks on the hot core. A possible interpretation is that this feature is  a CO component, since
high excitation methanol lines are unlikely to be detected at positions distant from hot cores,
 and that contamination with CH$_3$OH happens at the hot core position.

\citet{2000MNRAS.316..152M} show that the absorption at 6.5~km~s$^{-1}$ (see Figs.~\ref{red}-~\ref{blue})
is real, not coming from the off position used for position switching.

\section{Physical parameters of the outflow}

To determine the physical parameters in the outflow, we used the CO
observations and a spherically symmetric LVG
statistical equilibrium code, with the cosmic background as the only
radiation field.  To compare the observations at the three
different frequencies with the LVG results, we smoothed the CO(4-3)
and CO(7-6) data to the same spatial resolution of the CO(3-2), which is $18''$. To reduce the
noise level in the spectra,
we resampled them to a resolution of 5~${\rm km~s}^{-1}$.\\
The non-detection of the $^{13}$CO(3-2) transition in
the wings gives an upper limit to the CO column density of 10$^{17}$~cm$^{-2}$ for 
the $(-10'',-14'')$ and $(17.5'',14'')$ positions. In this calculation, we 
used the relation between $^{12}$C and $^{13}$C given by \citet{1994ARA&A..32..191W} and a galactocentric 
distance of 6.8~kpc \citep{1998ApJ...503..785K}.  
All models were
run for this value of the column densiy. At this column density, the CO transitions are optically thin, 
with the exception of CO(3-2), which is optically thick for
T$<50$~K, n(H$_2)<10^4$~cm$^{-3}$. Hence, our results should not
depend much on the column density used in the calculations. The code inputs are the
density and the kinetic temperature of the gas. Since a direct estimate of the
latter is not available, we ran models for several densities and
temperatures. 

Source sizes play an important role in comparing the observations to
the results of the LVG simulations, as the main beam temperature of
a transition is equal to $\frac{s^2}{s^2+\Omega^2}\times T_l$, where
$s$ and $\Omega$ are the source size and the beam size  in arcsec, and $T_l$ is the line
intensity from the LVG analysis. In our case, the beam is
the same for all the transitions, but the source sizes may differ as
the three lines have different excitation conditions. Therefore, 
line ratios may also be affected by the uncertainties in the beam filling
factors. In the following, we will analyse the line ratios between the
three transitions (upper J divided by lower J), assuming that all
transitions are emitted by the same volume of gas. These are lower
limits to the effective line ratios, as less excited transitions are
likely to come from more extended regions.

\begin{figure}[]
\centering
\includegraphics[angle=-90,width=6cm]{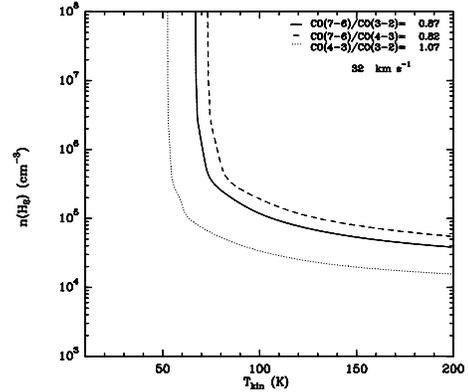}
\caption{Line ratios as a function of temperature and density 
toward the  $(17.5'',14'')$ position.}\label{red-lvg}
\end{figure}

In Fig.~\ref{red-lvg}, the three different line ratios at
$(17.5'',14'')$, $v_{lsr}=32{\rm~km~s}^{-1}$, are plotted as a
function of density and kinetic temperature. 
Since we have ignored the effects of beam dilution, which are likely to be important 
at least for the CO(7-6)/CO(3-2) line ratio,
we can only estimate lower limits
to the gas parameters. We find that the
temperature must be higher than 50~K, while the density is
of the order of $10^4$~cm$^{-3}$, equivalently a thermal pressure of
$5\times 10^5$ $\hbox {{\rm cm}}^{-3}$ K. Moreover, as the line ratios
between the CO(7-6) line and the other transitions show, there is also
a denser, warmer component, from which
the CO(7-6) line is emitted. As already discussed, no corrections for the different beam
filling factors are applied, and the values we used for the line ratios are
likely to be lower limits. Therefore, the values  of $n=7 \times 10^4$~cm$^{-3}$ and $T=70$~K
(corresponding to $P_{thermal}\sim 5\times 10^6$ $\hbox {{\rm cm}}^{-3}$~K) that we derived for the component
traced by CO(7-6) are lower limits \citep[consistent with other studies of high-J transitions in outflows, e.g. ][]{2002A&A...387..931B}. 
Taking into account a calibration error of 0.2 for each ratio, we find $30<T<55$~K, $7\times 10^3 <n< 2\times 10^4$~cm$^{-3}$. For the warmer component traced by CO(7-6)/CO(4-3) $55<T<95$~K, $3\times 10^4 <n< 1\times 10^5$~cm$^{-3}$.
The results for the other velocity channels 
are very similar, as the line ratios between the different CO
transitions do not  significantly change at this position
(Fig.~\ref{red}).

\begin{figure}[]
\centering
\subfigure[]{\includegraphics[angle=-90,width=6cm]{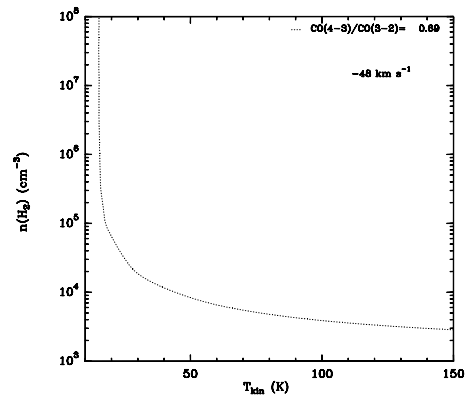}\label{1blue-lvg}}
\subfigure[]{\includegraphics[angle=-90,width=6cm]{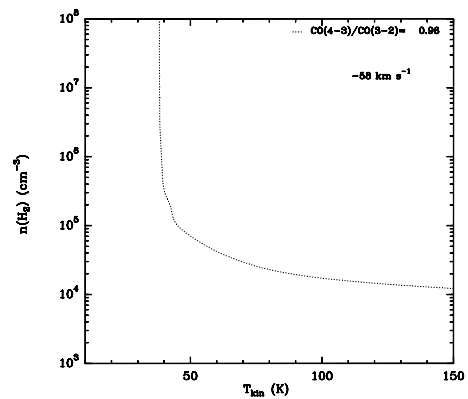}\label{3blue-lvg}}
\caption{CO(4-3)/CO(3-2) line ratio as a function of temperature and density 
for two velocity channels toward the  $(-10'',-14'')$ position.}\label{blue-lvg}
\end{figure}

For the analysis on the $(-10'',-14'')$ position, we only used the
CO(4-3)/CO(3-2) ratio, since the signal-to-noise ratio in the CO(7-6)
spectrum is not very high in the wings ($\sim 3$). The lower limits on the 
temperature and the density of the gas vary along the wings, ranging 
between 15 and 50~K, $2 \times 10^3$ and $2 \times 10^4$~cm$^{-3}$ (see Figs.~\ref{1blue-lvg}-~\ref{3blue-lvg}).
Taking into account the uncertainties on the calibration, the lower limits  we derive range 
between 10 and 30~K and $n<10^4$~cm$^{-3}$ for $v_{lsr}=-48{\rm~km~s}^{-1}$, and $20<T<56$~K and $4\times 10^3 <n< 3\times 10^4$~cm$^{-3}$ for $v_{lsr}=-58{\rm~km~s}^{-1}$.

The difference in temperatures and densities derived for the  red and blue lobes may 
be caused by the (pre-existing) density structure of the molecular cloud into which the outflow impinges 
and with which it interacts. Continuum maps of the region \citep{2000A&A...358..242S} support this scenario, since the continuum emission
is extended to the north of NGC6334~I.

The CO column density of the high velocity gas can be derived from the
integrated intensity, by assuming that the level populations are in
LTE at the kinetic temperature derived in our calculations (30~K for the blue
lobe, 50~K for the red one). To avoid
contamination from the core emission, we integrate between 10 and
59~${\rm km~s}^{-1}$ for the redshifted gas, and -88 to -22~${\rm
  km~s}^{-1}$ for the blueshifted gas, which results in lower limits to the
derived masses.  The momentum $p$, the energy
$E$, and the characteristic time scale $t$ can be derived assuming
that the outflow is a constant velocity flow with true velocity equal
to the maximum velocity observed
\citep[e.g.][]{1985ARA&A..23..267L,2002A&A...383..892B}. Finally, the
mass entrainment rate $\dot{M}_{tot}$, the mechanical force $F$, and
the mechanical luminosity $L$ can be derived from the other
parameters. Assuming that the inclination angle
between the outflow and the line of sight is $\theta$, the following relations
are used \citep[from][but corrected for the inclination angle $\theta$]{2002A&A...383..892B}:
 
\begin{equation}\label{colden}N(H_2)=10^4\frac{6.97\times10^{15}}{\nu^2\mu^2}T_{ex}e^{(E_u/T_{ex})}\int{T_{CO(4-3)}dv}
\end{equation}
\begin{equation}\label{mass}M_{tot}=\frac{1}{cos\theta}\left(N_b\times area_b+N_r\times area_r\right)m_{\rm H_2}
\end{equation}
\begin{equation}\label{momentum}p=\frac{1}{cos\theta}\left(M_b\times
    v_{max_b}+M_r\times v_{max_r}\right)
\end{equation}
\begin{equation}\label{energy}E=\frac{1}{cos^2\theta}\left(\frac{1}{2}M_b\times v_{max_b}^2+\frac{1}{2}M_r\times v_{max_r}^2\right)
\end{equation}
\begin{equation}\label{time}t=\frac{r}{2 sin\theta}\times \frac{cos\theta}{v_{max_b}+v_{max_r}}
\end{equation}
\begin{equation}\label{force}F=\frac{p}{t}
\end{equation}
\begin{equation}\label{mass rate}\dot{M}_{tot}=\frac{M_{tot}}{t}
\end{equation}
\begin{equation}\label{lum}L=\frac{E}{t}
\end{equation}

\begin{table}[h!]
\caption{Physical parameters of the outflow}\label{outflow}
\begin{tabular}{lccc}
\hline\hline
$N_b$&\multicolumn{3}{c}{$1.9\times 10^{21}$~cm$^{-2}$}\\
$N_r$&\multicolumn{3}{c}{$1.5\times 10^{21}$~cm$^{-2}$}\\
&\multicolumn{1}{c}{no corr.}&\multicolumn{1}{c}{$\theta=45^\circ$}&\multicolumn{1}{c}{$\theta=70^\circ$}\\
$M_b~[M_\odot]$&1.1&1.5&3.2\\
$M_r~[M_\odot]$&0.8&1.2&2.4\\
$M_{tot}~[M_\odot]^a$&1.9&2.7&5.6\\
$t~[10^3~{\rm yr}]$&4.4&4.5&1.6\\
$p~[M_\odot$~${\rm km~s}^{-1}]^b$&140&200&410\\
$E$~$[10^{47}$~erg]&1.0&2.1&8.9\\
$F_m ~[M_\odot~{\rm km~s}^{-1}~{\rm yr}^{-1}]$&0.03&0.04&0.3\\
$L_m~[L_\odot]$&780&1550&1800\\
$\dot{M}_{tot}~[10^{-3}~M_\odot~{\rm yr}^{-1}]$&1.9&2.6&15\\
\hline
\hline
\end{tabular}
\begin{list}{}{}
\item $^a$ \citet{1990A&A...239..276B}: 2.3~$M_\odot$; \citet{2000MNRAS.316..152M}: 4.8~$M_\odot$.
\item $^b$ \citet{1990A&A...239..276B}: 92~$M_\odot$~${\rm km~s}^{-1}$; \citet{2000MNRAS.316..152M}: 360~$M_\odot$~${\rm km~s}^{-1}$.

\end{list}
\end{table}

Estimates of the physical parameters in the outflow from NGC6334~I
have already been derived from the analysis of CO(2-1) and CO(3-2)
observations. Both \citet{1990A&A...239..276B} and \citet{2000MNRAS.316..152M} do not apply any
corrections for the inclination angle to their values. In Table~\ref{outflow}, the parameters are given for
no corrections for the inclination angle and for $\theta=45^\circ
\rm{and}~70^\circ$, as
the small overlap between the redshifted and the blue emission and the emission at high velocity 
indicate that 
$\theta$ is closer to $90^\circ$ than to $0^\circ$. For comparison, results
from  \citet{1990A&A...239..276B} and \citet{2000MNRAS.316..152M} are also
given.
Our values are in good agreement with
previous studies; however, the mass we derive, and therefore other
parameters also, 
are smaller than what
both \citet{1990A&A...239..276B} and \citet{2000MNRAS.316..152M} had derived,
due to the higher kinetic temperature we use for the red lobe. Morever, high-J transitions
trace only the warm gas, while low-J lines also trace a colder component.
Both this study
and \citet{2000MNRAS.316..152M} make use of the maximum observed velocity in
the outflow to derive the energetics of the outflow, while
\citet{1990A&A...239..276B} use the mean velocity. The estimate of
the mass is also affected by other uncertainties; the observed area of the
outflow depends on the transitions used in the calculations and the excitation temperature
used in Eq.~(\ref{colden}) has been derived for one position per lobe. \citet{2003MNRAS.339.1011G} 
also found lower outflow masses
in CO(3-2) than in CO(1-0); \citet {1990ApJ...348..530C}
have shown that masses derived using  Eq.~(\ref{mass}) are upper limits to
their true value.

\section{Conclusions}

We report observations in high excitation CO lines towards the high
velocity outflow from NGC6334~I. By using an LVG analysis, we have constrained
the temperatures to be higher than 50~K and the H$_2$ density to values higher
than $n\sim 10^4$~cm$^{-3}$ towards the peak
position in the red lobe, $T>15$~K and $n>10^3$~cm$^{-3}$ towards the peak
position in the blue lobe. The physical parameters of the outflow, its mass
and its energetics, have been computed using the temperatures derived from this
analysis.
Often \citep[e.g.][]{2002A&A...383..892B,kim}, low temperatures are assumed
in computing outflow masses; our analysis has shown that high kinetic
temperatures are present, and traced by high excitation CO lines, while 
they are not found when only low energy transitions are available. High spatial
resolution maps of CO transitions at high-J are needed to infer reliable
values of the kinetic temperatures and of the other physical parameters in the outflow.    

\acknowledgements{We would like to thank the anonymous referee for the pront reply. BP is grateful to the Alexander von Humboldt Foundation for a  Research Fellowship.} 
\bibliographystyle{aa} \bibliography{5338}
\end{document}